# Harnessing Machine Learning for Quantum-Accurate Predictions of Non-Equilibrium Behavior in 2D Materials

Yue Zhang[1#], Robert J. Appleton[2#], Kui Lin[3#], Megan J. McCarthy[4], Jeffrey T. Paci[5], Subramanian K.R.S. Sankaranarayanan[6], Alejandro Strachan[2], Horacio D. Espinosa[*1,7]

[1] Department of Mechanical Engineering, Northwestern University, Evanston, IL 60208, USA.
[2] School of Materials Engineering, Purdue University, West Lafayette, IN 47907, USA.
[3] Department of Civil and Environmental Engineering, The Hong Kong Polytechnic University, Hong Kong 999077, China.
[4] Center for Computing Research, Computational Multiscale Department, Sandia National Laboratories, Albuquerque, NM 87185, USA.
[5] Department of Chemistry, University of Victoria, Victoria, British Columbia V8W 3V6, Canada.
[6] Center for Nanoscale Materials, Argonne National Laboratory, Lemont, IL 60439, USA.
[7] Theoretical and Applied Mechanics Program, Northwestern University, Evanston, IL 60208, USA.

[#]These authors contributed equally

E-mail: espinosa@northwestern.edu; strachan@purdue.edu

**Abstract:** Accurately predicting the non-equilibrium mechanical properties of two-dimensional (2D) materials is essential for understanding their deformation, thermo-mechanical properties, and failure mechanisms. In this study, we parameterize and evaluate two machine learning (ML) interatomic potentials, SNAP and Allegro, for modeling the non-equilibrium behavior of monolayer $MoSe_2$. Using a density functional theory (DFT) derived dataset, we systematically compare their accuracy and transferability against the physics-based Tersoff force field. Our results show that SNAP and Allegro significantly outperform Tersoff, achieving near-DFT accuracy while maintaining computational efficiency. Allegro surpasses SNAP in both accuracy and efficiency due to its advanced neural network architecture. Both ML potentials demonstrate strong transferability, accurately predicting out-of-sample properties such as surface stability, inversion domain formation, and fracture toughness. Unlike Tersoff, SNAP and Allegro reliably model temperature-dependent edge stabilities and phase transformation pathways, aligning closely with DFT benchmarks. Notably, their fracture toughness predictions closely match experimental measurements, reinforcing their suitability for large-scale simulations of mechanical failure in 2D materials. This study establishes ML-based force fields as a powerful alternative to traditional potentials for modeling non-equilibrium mechanical properties in 2D materials.

# Introduction

Transition metal dichalcogenides (TMDs) have attracted significant interest due to their exceptional electronic, optical, thermal, and mechanical properties[1-6]. These characteristics make them promising for applications in flexible electronics, wearable sensors, thermoelectric devices, and have the potential to enable further scaling of CMOS technology while reducing power[7]. However, such applications often involve dynamic, non-equilibrium mechanical deformations—including stretching, bending, and twisting—that directly affect device performance and reliability. Designing and optimizing TMDs-based technologies thus requires a detailed atomic-scale understanding of their non-equilibrium mechanical properties.

Despite advancements in first-principles calculations and experimental techniques, a comprehensive understanding of TMDs' ultimate mechanical behavior under non-equilibrium conditions remains challenging. Density functional theory (DFT) provides accurate predictions, see, for example, Ref.[8], but is computationally impractical for large-scale simulations. Experimental methods, meanwhile, are limited by sample quality, measurement resolution, and the complexity of simulating non-equilibrium conditions[9,10]. Molecular dynamics (MD) simulations based on interatomic potentials offer a viable alternative by bridging atomic-scale interactions with macroscopic properties, but their accuracy hinges on the choice of interatomic potential. Traditional force fields, such as the Tersoff potential[11] and ReaxFF[12,13], use expressions based on chemical and physical intuition and often struggle to accurately capture complex bonding and structural behaviors, particularly under large deformations or in the presence of defects. Recent work by Espinosa and co-workers further highlighted accuracy trade-offs in classical force fields when attempting to predict a broad range of material properties simultaneously[14,15]. To overcome these limitations, machine learning (ML) interatomic potentials have emerged as a powerful solution[16-19], achieving near DFT accuracy while maintaining computational efficiency suitable for large-scale simulations. A key question about ML potentials is their ability to extrapolate to configurations not seen during training. Probing extrapolation capabilities of ML potentials is routinely done by running dynamical simulations that explore phases outside of the training configurations[20] (i.e., melting temperature simulations) but has also been measured more rigorously using geometric criteria that compare the feature space captured within the training data to configurations in the test data[21].

In this study, we evaluate the ability of two ML interatomic potentials Spectral Neighbor Analysis Potential (SNAP)[22] and Allegro[23] to learn atomic interactions associated with non-equilibrium configurations of $MoSe_2$ and, importantly, to generalize to new configurations. SNAP, a descriptor-based potential, represents many-body atomic environments using bi-spectrum components and integrates seamlessly with LAMMPS[24]. Allegro, leveraging a deep equivariant neural network architecture, achieves state-of-the-art accuracy with a local formulation, making it particularly suitable for modeling non-equilibrium properties in large systems. The selection of SNAP and Allegro is motivated by their complementary strengths: SNAP offers robustness and interpretability, particularly in systems with well-defined symmetry, while Allegro provides higher flexibility and accuracy in capturing nonlinear atomic interactions, defects, and phase transitions. By systematically assessing stress-strain behavior, phase transition energetics, edge stability, and fracture toughness of monolayer $MoSe_2$, we aim to benchmark SNAP and Allegro against both DFT data and classical potentials, e.g., Tersoff. The study provides a quantitative framework for evaluating ML-based force fields' accuracy, cost, and suitability for large-scale simulations of mechanical properties in 2D materials.

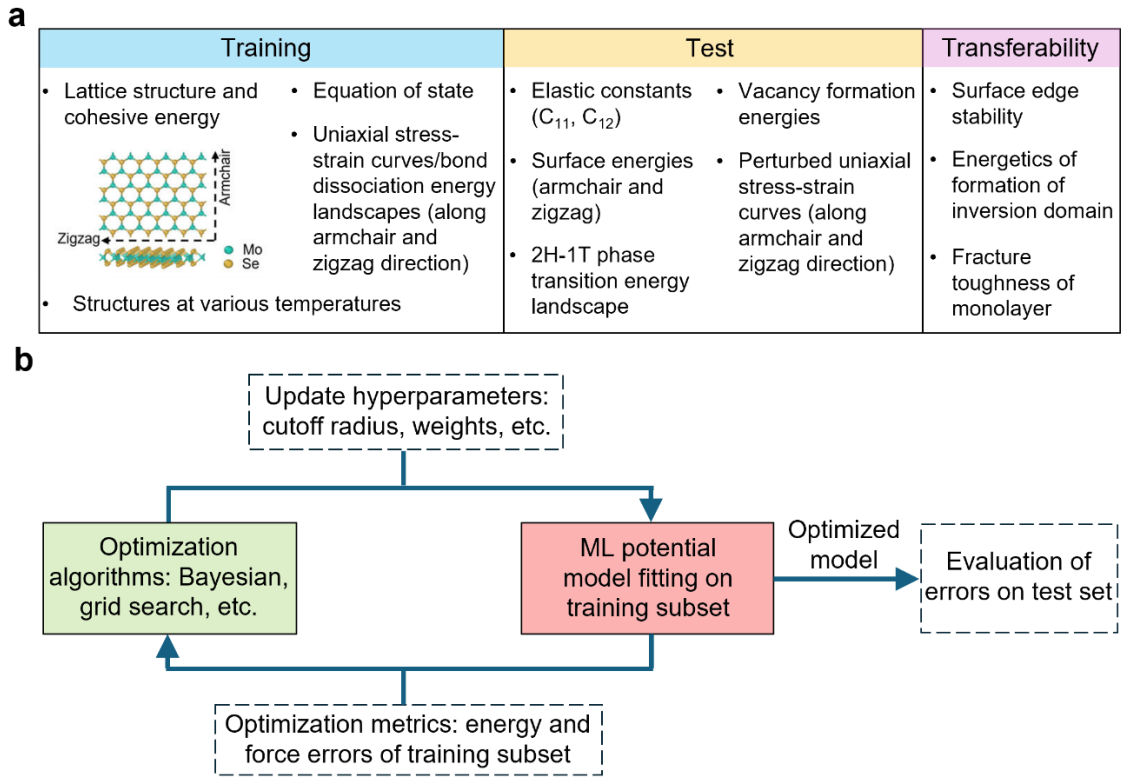

Fig. 1 | (a) Overview of the properties included in the training and test sets of the database and the properties used to study the transferability. (b) Framework for parameterizing ML potentials.

## Results

Figure 1a summarizes the properties included in the training and test datasets used for parameterizing the ML interatomic potentials, SNAP and Allegro. The overall framework for ML potential parameterization is outlined in Fig. 1b, with methodological details described in the **Methods** section. After training, the performance of the potentials is assessed by evaluating their errors on the test dataset. The ML potentials that achieve the lowest overall errors are selected for further comparisons against ab initio calculations and the Tersoff force field, as well as for investigating their transferability in predicting properties beyond the training and test sets (Fig. 1a). Figs. 2 and S1 compare the performance of parameterized ML interatomic potentials (SNAP and Allegro) against ab initio DFT calculations and the Tersoff potential for monolayer $MoSe_2$. The results highlight the limitations of the Tersoff potential in capturing non-equilibrium properties and defects in $MoSe_2$, whereas ML-based potentials provide significantly improved accuracy.

Figures. 2a and 2b show uniaxial stress–strain curves along the zigzag direction for pristine $MoSe_2$ and $MoSe_2$ with a 4|4E grain boundary. The Tersoff potential deviates markedly from ab initio results at strains beyond 5% for the defective structure, while SNAP and Allegro accurately capture the stress–strain response across the entire deformation process, including non-equilibrium stages.

Fig. 2c demonstrates that both ML and Tersoff potentials perform well for equilibrium properties such as the equation of state for pristine $MoSe_2$. However, Fig. 2d reveals that the Tersoff potential significantly deviates from ab initio results for reaction coordinates beyond 0.4 in the square phase transition energy landscape. Moreover, it incorrectly predicts the final phase as the minimum energy state, underscoring its limitations in modeling phase transitions. In contrast, SNAP and Allegro accurately reproduce the entire transition process, correctly identifying the initial phase as the most stable state.

Figs. 2e and 2f depict bond dissociation energy landscapes along the armchair direction for pristine $MoSe_2$ and zigzag for $MoSe_2$ with a 4|4E grain boundary. The Tersoff potential diverges

significantly from ab initio calculations when the distance change exceeds 10 Å in the defective structure, whereas SNAP and Allegro maintain accuracy throughout the deformation process.

Overall, these results demonstrate the superior flexibility of the ML potentials in capturing a broader range of atomic environments compared to traditional force fields. The parameterized SNAP and Allegro potentials more accurately reproduce stress–strain relations, bond dissociation energies, and phase transition energies from the training dataset, offering a more precise description of interatomic interactions.

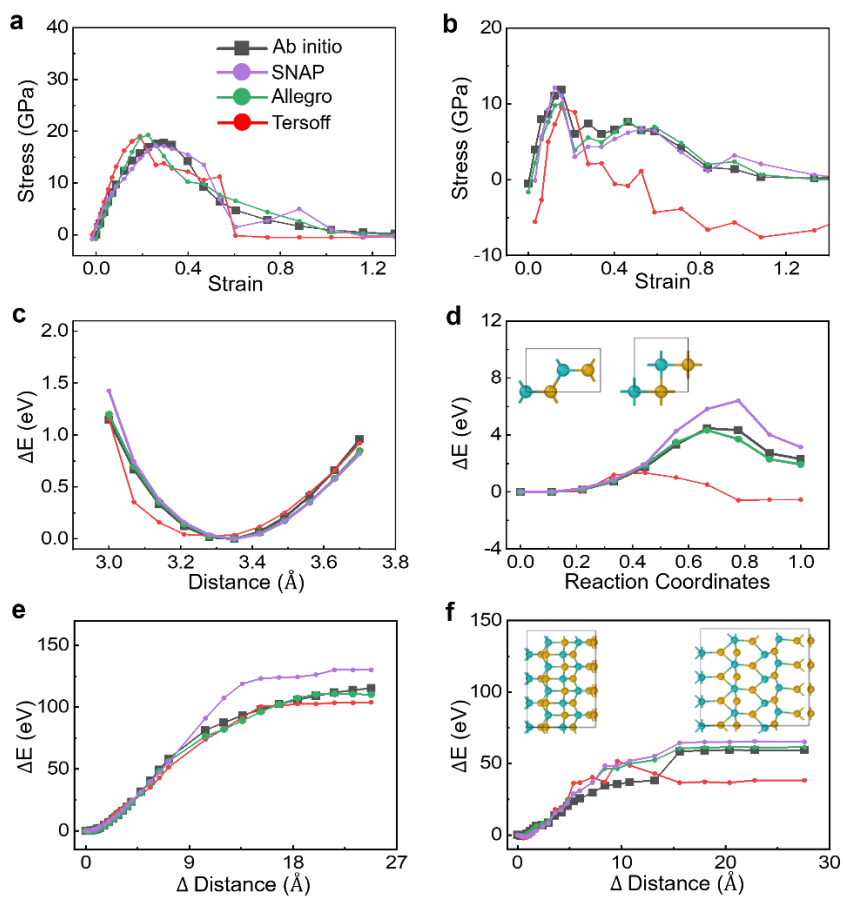

Fig. 2 | Material processes in the training set using Tersoff, SNAP, and Allegro potentials, compared to ab initio results. (a) Uniaxial stress–strain curve for pristine $MoSe_2$ along the zigzag direction. (b) Uniaxial stress–strain curve for $MoSe_2$ with 4|4E grain boundaries along the zigzag direction. (c) Equation of state. (d) Square phase transition energy landscape, with snapshots corresponding to reaction coordinates of 0 and 1. Se atoms are shown in yellow, and Mo atoms in cyan. (e) Bond dissociation energy landscape along the armchair direction. (f) Bond dissociation energy landscape for $MoSe_2$ with 4|4E grain boundaries along the zigzag direction, with snapshots corresponding to Δ Distance of 0 and 5.4 Å, respectively.

Figure 3 and Table 1 compare the predictions of material properties in the test set using parameterized ML interatomic potentials (SNAP and Allegro) against ab initio DFT calculations and the traditional Tersoff potential. The results demonstrate that ML-based potentials accurately capture properties beyond the training dataset, whereas the Tersoff potential exhibits limited accuracy. Figs. 3a and 3b present perturbed stress–strain curves along the zigzag and armchair directions. Perturbed stresses refer to the soft modes, where phonon instabilities are considered[14]. The Tersoff potential deviates significantly from ab initio results at large strains, while SNAP and Allegro accurately capture the stress–strain response throughout the entire deformation process, including non-equilibrium stages. Fig. 3c shows the phase transition energy landscape from the 2H to 1T phase. The Tersoff potential deviates from ab initio results at reaction coordinates beyond 0.5 and incorrectly predicts the 1T phase energy to be close to that of the 2H phase. In contrast, SNAP and Allegro accurately reproduce the transition energy profile, correctly identifying the 2H phase as the more stable state.

Table 1 compares vacancy formation energies (seven vacancy types) and surface energies (armchair and zigzag) predicted by each potential. We stress that the training does not contain these properties. The Tersoff potential shows significant deviations from ab initio results, whereas ML-based potentials provide far more accurate predictions. These findings highlight the superior accuracy of ML potentials for predicting properties beyond the training set. This enhanced predictive power enables ML potentials to model new phenomena and conditions outside the initial parameterization scope. We quantified the degree of extrapolation in terms of the local structures sampled in the test vs. training set. We used a dimensionality reduction technique, uniform manifold approximation and projection (UMAP),[25] to visually compare the atomic environments sampled in the training and test sets of the Allegro potential. As described in Fig. S2, we found that several local structures in the test set had never been seen by the potential. Not surprisingly, this is most noticeable in the structures containing vacancies. Furthermore, Fig. S3 shows that the simulations in the transferability set also explored areas of the feature space not captured in the training data.

While ML potentials are known for their higher accuracy, an important issue is to assess the increase in computational cost and its impact on the feasibility of large-scale atomistic simulations. We benchmarked the trade-off between computational cost, CPU time per MD step per atom, and the root mean square error (RMSE) of the test set for increasingly more complex descriptions of the ML potentials, as shown in Fig. 3d. Namely, the parameter $J_{max}$ for SNAP and $l_{max}$ for Allegro. The computational costs are based on single-CPU core calculations in LAMMPS for MoSe$_2$. The models with the lowest RMSE were used for accuracy comparisons. The results show that ML potentials achieve significantly higher accuracy than the traditional Tersoff force field while maintaining comparable computational efficiency in molecular dynamics simulations. Among ML potentials, Allegro offers the best trade-off, delivering DFT-level accuracy with lower computational cost than SNAP due to its advanced deep-learning architecture and local equivariant representation. Moreover, the speed of Allegro potential ($l_{max}$=3) is five orders of magnitude faster than DFT-MD for the same simulation system with 672 atoms based on single-CPU core, and this difference of speed will further increase with system size.

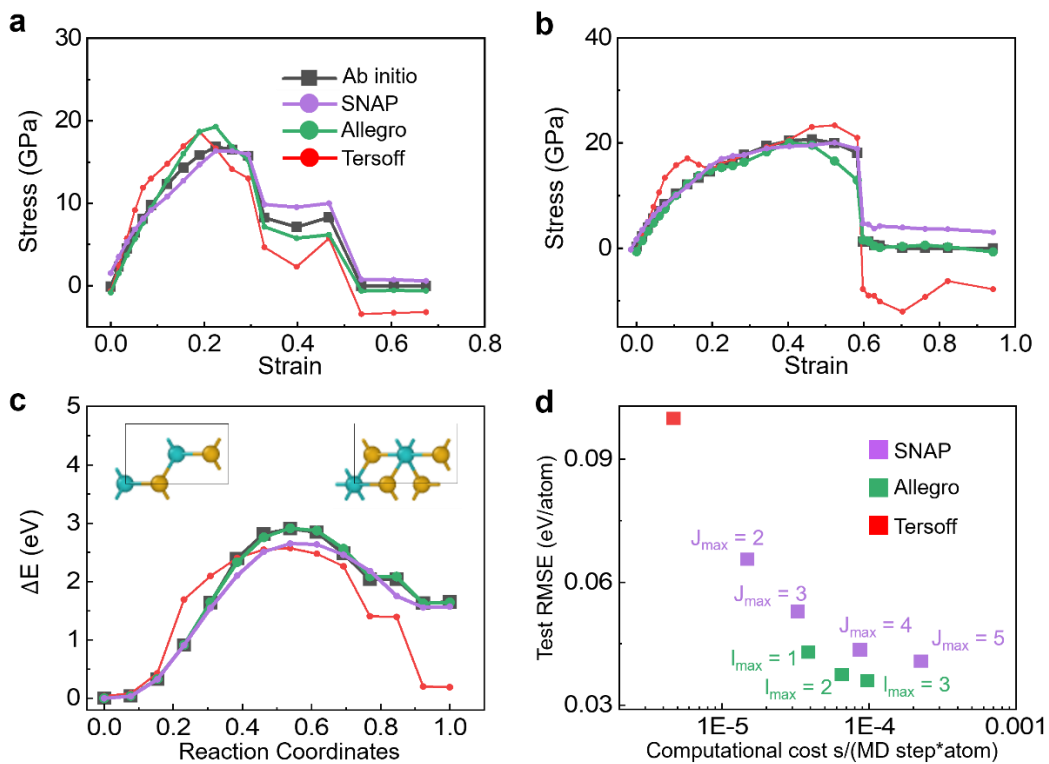

Fig. 3 | Predictions of material properties in the test set using parameterized potentials, compared to ab initio results. (a) Perturbed uniaxial stress–strain relation along the zigzag direction. (b) Perturbed uniaxial stress–strain relation along the armchair direction. (c) 2H–1T phase transition energy landscape, with snapshots corresponding to reaction coordinates of 0 and 1. (d) Trade-off between errors and computational costs for various potentials.

To further evaluate the transferability of interatomic potentials, we examined the ability of parameterized ML potentials to predict the surface edge stability of monolayer MoSe$_2$ at different

temperatures—a key property for nanoelectronics and catalysis that was not explicitly included in the training dataset. Table 2 presents the stability predictions of various edge types at 300 K and 650 K using SNAP, Allegro, and Tersoff potentials, compared to ab initio molecular dynamics (AIMD) results. We examined four edge types, Mo-Klein, Mo-zigzag, Se-zigzag, and Armchair, whose atomic structures are depicted in Fig. S4.

Three distinct stability behaviors were observed: (i) the edge undergoes relaxation maintaining the same bonding environments of the initial structure (Fig. S4a), labeled as *relaxation* in Table 2; (ii) the edge becomes disordered (Fig. S4b, d), labeled as *disordered reconstruction*; and (iii) the edge undergoes reconstruction into another ordered structure (Fig. S4c), labeled as *ordered reconstruction*.

|  | Ab initio | SNAP | Allegro | Tersoff |
|---|---|---|---|---|
| $C_{11}$ (GPa) | 129.34 | 135.1 | 140 | 139.39 |
| $C_{12}$ (GPa) | 35.36 | 31.71 | 34 | 29.44 |
| $\Gamma_{AC}$ (eV/Å) | 0.77 | 0.69 | 0.73 | 0.50 |
| $\Gamma_{ZZ}$ (eV/Å) | 0.72 | 0.58 | 0.69 | 0.44 |
| $E_{Se}$ (eV) | 3.30 | 3.56 | 3.15 | 2.81 |
| $E_{Se2}$ (eV) | 6.23 | 7.35 | 5.68 | 5.55 |
| $E_{MoSe3}$ (eV) | 10.94 | 9.98 | 9.79 | 9.25 |
| $E_{MoSe6}$ (eV) | 20.27 | 19.34 | 17.43 | 14.38 |
| $E_{Mo}$ (eV) | 6.51 | 4.75 | 3.87 | 3.56 |
| $E_{Mo2F}$ (eV) | 11.69 | 6.32 | 7.73 | 7.05 |
| $E_{Mo2C}$ (eV) | 10.44 | 5.95 | 5.35 | 6.64 |

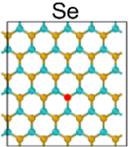
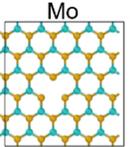
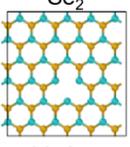
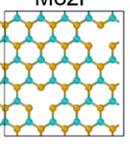
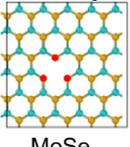
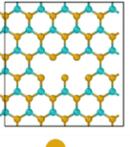
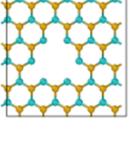
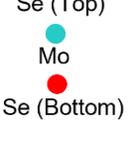

Table 1 | Comparison of property predictions in the test set using parameterized potentials and ab initio results. The comparison includes elastic constants ($C_{11}$ and $C_{12}$), surface energies along the armchair ($\Gamma_{AC}$) and zigzag ($\Gamma_{ZZ}$) directions, and vacancy formation energies for various defect types: Mo monovacancy ($E_{Mo}$), non-adjacent Mo divacancies ($E_{Mo2F}$), adjacent Mo divacancies ($E_{Mo2C}$), Se monovacancy ($E_{Se}$), Se divacancies with atoms above and below the Mo layer ($E_{Se2}$), one Mo and three adjacent Se vacancies within the same Se atomic layer ($E_{MoSe3}$), and one Mo with six adjacent Se vacancies ($E_{MoSe6}$). The images on the right of the table show the atomic configurations of various vacancies.

The simulation results in Table 2 indicate that Allegro successfully reproduces AIMD predictions for all edge types at both 300 K and 650 K, while SNAP and Tersoff fail to capture the surface reconstruction of the Mo-zigzag edge at 300 K. Additionally, Tersoff over-stabilizes the Mo-zigzag edge at 650 K.

To understand the varying performance of these potentials for the Mo-zigzag edge at 300 K, we calculated the energy difference between the initial and reconstructed surface structures using these potentials and compared them to DFT results (Table S1). The results show that Allegro correctly predicts a lower energy for the reconstructed surface, in agreement with DFT, indicating its stability. In contrast, SNAP and Tersoff predict an opposite energy difference, explaining their failure to reproduce surface reconstruction in MD simulations.

| Edge type | Mo-Klein | | Mo-zigzag | | Se-zigzag | | Armchair |
|---|---|---|---|---|---|---|---|
| Temperature | 300K | 650K | 300K | 650K | 300K | 650K | 300K |
| AIMD | relaxation | disordered reconstruction | ordered reconstruction | disordered reconstruction | relaxation | relaxation | disordered reconstruction |
| SNAP | relaxation | disordered reconstruction | relaxation | disordered reconstruction | relaxation | relaxation | disordered reconstruction |
| Allegro | relaxation | disordered reconstruction | ordered reconstruction | disordered reconstruction | relaxation | relaxation | disordered reconstruction |
| Tersoff | relaxation | disordered reconstruction | relaxation | relaxation | relaxation | relaxation | disordered reconstruction |

Table 2 | Transferability test of parameterized potentials for edge stability of monolayer MoSe$_2$. The same edge configurations were equilibrated using ab initio molecular dynamics (AIMD) for 1 ps and classical molecular dynamics (MD) for 500 ps at the specified temperature. The *disordered reconstruction* category refers to cases where the surface structure becomes disordered (Fig. S4), while the *ordered reconstruction* category denotes cases where the surface transitions from its initial structure to another stable configuration.

In another transferability test, we examined the ability of the parameterized ML potentials to predict the formation of inversion domains in monolayer MoSe$_2$. High-resolution *in situ* TEM experiments revealed the formation of such inversion domains, which occurs in two steps, formation of a line vacancy resulting in a 4|4E GB-like structure (step 1) and transformation of the 4|4E GB-like structure to an inversion domain (step 2) assisted by additional migration of vacancies[26].

DFT optimization and AIMD simulations provided no indication that a single line vacancy could evolve into a 4|4E GB-like structure. Instead, a two-line vacancy model (Fig. 4a), spontaneously transformed into a 4|4E GB-like structure (Fig. 4b) upon relaxation. Moreover,

AIMD calculations confirmed the stability of this structure at 300 K. The energy difference between the initial two-line Se-line vacancy configuration and the final 4|4E configuration was found to be -45.4 meV/atom, Fig. 4c. Likewise, the SNAP, Allegro, and one parameterization of Tersoff, successfully reproduce energy reduction during the formation of the 4|4E GB-like structure, although with various degrees of accuracy. MD simulations at 300 K corroborate these findings, showing the transformation of two-line vacancies into the 4|4E GB-like structure with a lower energy state. Another parameterization of the Tersoff potential, optimized for fracture[27], predicts a higher energy state for the 4|4E GB-like structure.

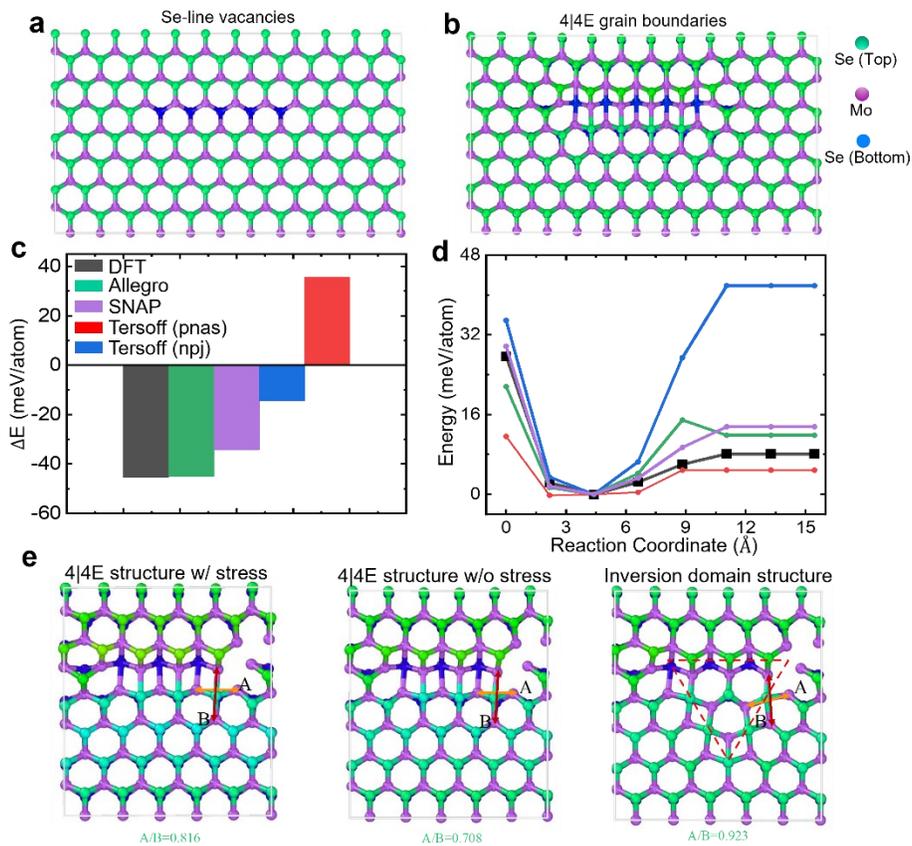

Fig. 4 | (a) Atomic configuration of Se-line vacancies. (b) Atomic configuration of 4|4E grain boundaries. (c) Energy difference $\Delta E$ between the 4|4E grain boundaries configuration and the Se-line vacancy configuration, as calculated using the parameterized potentials and compared to DFT results. (d) Energy profile for the transition from the 4|4E grain boundary structure to the inversion domain, calculated using parameterized potentials and compared to DFT results. (e) Atomic

configurations of 4|4E grain boundaries with and without applied stress, along with the inversion domain structure.

According to Lin et al.,[26] during step 2, Se vacancies are generated either by irradiation damage or migration at one end of the 4|4E GB-like structure nucleate formation of the inversion domain. Experimental observations suggest that this process involves structural shrinkage, characterized by changes in the A/B ratio of the Mo sublattice, as indicated in Fig. 4e. Accordingly, we constructed initial and final structures closely resembling those observed experimentally (Fig. 4e) and performed transition path simulations. DFT calculations indicate that the initial structure is under stress (A/B = 0.816), leading to significant cell distortion between the initial and final states. To obtain the energetics along the path between the initial and final atomic configurations, we employed the solid-state NEB (SSNEB) method, Fig. 4d. The energy profile reveals that the system transitions from the initial configuration to the lowest-energy intermediate state. Further analysis shows that this intermediate state retains the 4|4E GB-like structural characteristics and has an A/B ratio of ~0.702, closely matching the stress-free 4|4E GB-like structure (A/B ≈ 0.708) obtained from DFT calculations. Beyond this intermediate state, the energy increases slightly, with the final structure exhibiting an energy level of 8.1 meV/atom higher than the lowest-energy intermediate state. Fig. 4d also shows that SNAP, Allegro, and one of the parameterized Tersoff, successfully reproduce this energy profile during inversion domain formation, whereas another parameterization of Tersoff significantly overpredicts the inversion domain energy.

These findings suggest that the nucleation process does not encounter a significant energy barrier but instead progresses through a stable, low-energy intermediate state. This intermediate configuration likely plays a crucial role in the experimental nucleation process by providing a natural pathway for stress relaxation and structural reorganization. Furthermore, the results emphasize that, due to the interplay between stress, vacancy distribution, and lattice evolution, multiple transitional configurations may exist between locally stable states. This does not exclude the possibility of a more intricate mechanism in real materials. Future research, including direct experimental observations and Monte Carlo simulations, will be valuable in further elucidating the kinetics and microstructural evolution underlying formation of inversion domains.

Finally, we evaluated the transferability of the parameterized ML potentials by comparing fracture toughness predictions on MoSe$_2$ to *in situ* fracture measurements using high-resolution electron microscopy (HRTEM)[27]. MD fracture simulations were performed on MoSe$_2$ monolayers consisting of a rectangular domain containing a notch (obtained by removing 5 atomic layers) and stretched along the zigzag direction, Fig. 5a. Free boundary conditions are imposed on edges normal to the notch and periodic boundaries are used in the direction of deformation. In the experiments, monolayers were not completely free in the out-of-plane direction but restrained by adhesion with the TEM grids. Hence, to mimic the experimental condition, we added potential surfaces, 20 Å away from the monolayer, to prevent large out-of-plane motions. The calculation of the J-integral, a fracture parameter introduced by J. Rice, was performed using the method reported in Xu et al.[27] Based on the results of MD simulations, we first mapped the displacement field from atomic positions using a cylindrical kernel function (with an averaging radius of 3 Å) and calculated the strain field based on a numerical gradient approximation. The stress field is mapped independently based on the per-atom virial stress output. We note that this stress only carries an approximate meaning in the case of many-body interatomic potentials, like Tersoff. This stems from the equal distribution of many body terms among the contributing atoms, leading to a non-conserving field. To achieve a more accurate mapping, central constraints must be applied. However, the stress field mapped on the per-atom virial stress output was shown to achieve a continuum stress within 5% of the exact value.

The stress fields $\sigma_{yy}$ and $\sigma_{xy}$, obtained from MD simulations using SNAP, Tersoff[27] and Allegro potentials, are shown in Fig. 5 c, d (SNAP), e, f (Tersoff), g and h (Allegro). The corresponding potential energies are depicted in Fig. S5. The critical energy release rates J, calculated based on these stress fields, are plotted in Fig. 5b, including the value measured using *in situ* HRTEM experiments[27]. SNAP, Allegro and Tersoff (pnas)[27], predict fracture toughnesses in good agreement with the average experimental measurement. In contrast, another parameterization of the Tersoff (npj) potential[14] underpredicts toughness. The fracture toughness calculated using SNAP with different values of $J_{max}$ are reported in Table S2, which indicates that the accuracy of calculated fracture toughness increases with the increase in the number of bi-spectrum components, $J_{max}$. Similarly, the fracture toughness calculated using Allegro with different values of $l_{max}$ are also reported in Table S2, which shows even the lowest value ($l_{max}$=1) matches reasonably well with experiment.

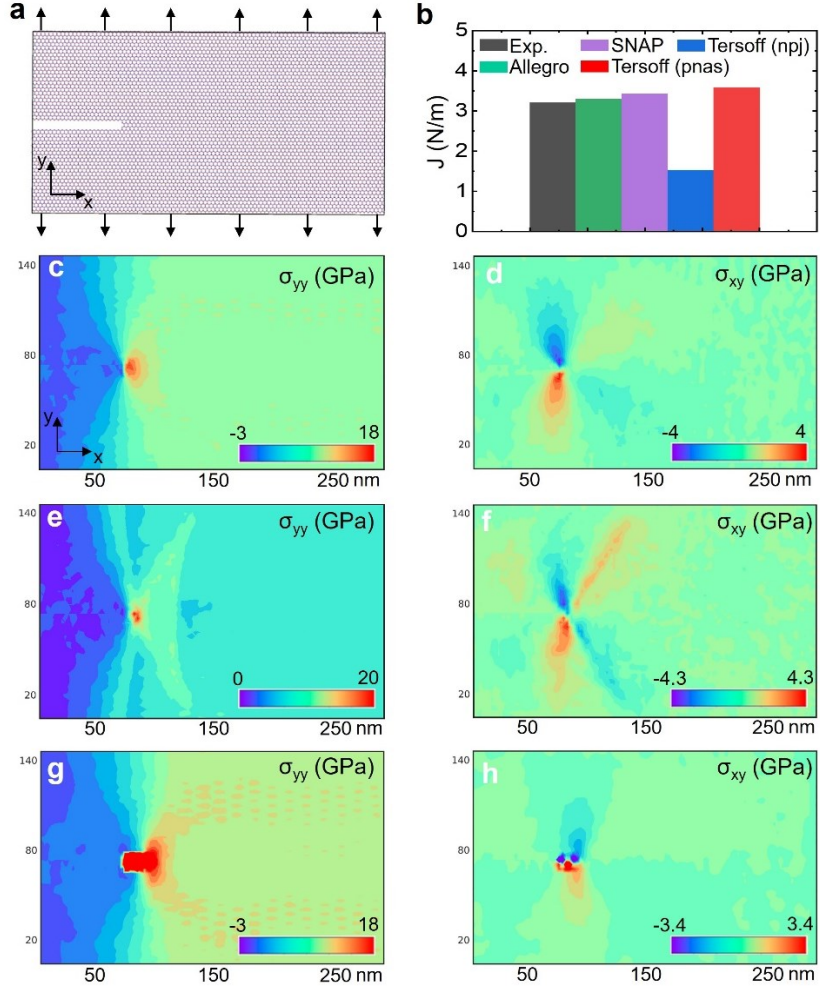

Fig. 5 | (a) The configuration used in MD simulations of fracture in monolayer MoSe$_2$. (b) Fracture toughness calculated using parameterized potentials in comparison to experimental measurement. Cauchy stress (c) $\sigma_{yy}$ and (d) $\sigma_{xy}$ in MoSe$_2$ calculated with the SNAP potential. Cauchy stress (e) $\sigma_{yy}$ and (f) $\sigma_{xy}$ calculated with the Tersoff potential. Cauchy stress (g) $\sigma_{yy}$ and (h) $\sigma_{xy}$ calculated with the Allegro potential.

## Discussion

In this work, we parameterized and evaluated machine learning (ML) interatomic models, specifically SNAP and Allegro, for predicting the non-equilibrium properties of monolayer MoSe$_2$. By comparing their predictions with density functional theory (DFT) calculations and the classical Tersoff model, we demonstrated that the ML models significantly enhance accuracy in capturing stress-strain relations, bond dissociation energies, phase transition energetics, and defect-related properties. These results highlight the ability of ML models to represent complex atomic environments more effectively than classical force fields, particularly in large deformation regimes and defective materials. Notably, SNAP and Allegro outperform Tersoff in both training and test datasets, demonstrating superior flexibility in capturing out-of-sample properties, including vacancy formation energies, surface stability, and phase transitions. Furthermore, Allegro achieves higher accuracy than SNAP at the same computational cost due to its advanced neural network architecture.

Beyond accuracy, ML models exhibit strong transferability, effectively predicting properties observed in HRTEM experiments well beyond their training sets. These include surface edge stabilities with various terminations, inversion domain formation, and fracture toughness. Both SNAP and Allegro accurately reproduce temperature-dependent edge stabilities and the energetics of inversion domain formation, underscoring their robustness in studying defects and defect engineering in 2D materials. Additionally, their fracture toughness predictions closely match experimental measurements, validating ML models for large-scale simulations of material deformation and failure.

Overall, this study establishes ML-based interatomic models as powerful tools for accurately modeling non-equilibrium mechanical properties in MoSe$_2$, providing a viable alternative to conventional force fields with DFT-level accuracy at modest computational costs. The success of SNAP and Allegro suggests broader applicability to other 2D materials. To explore this generalization, future work should focus on expanding training datasets, assessing temperature-dependent properties, and integrating ML models into multi-scale frameworks to bridge atomic-scale simulations with real-world applications. These advancements will enable more accurate, reliable, and computationally efficient simulations of non-equilibrium mechanical phenomena in emerging 2D materials.

## Methods

### DFT and ab initio MD calculation methods

In this work, the DFT calculations were performed with the SIESTA 4.1.5 software[28-30]. Nonrelativistic norm-conserving Troullier-Martins pseudopotentials[31] were used for the molybdenum and selenium atoms. The generalized gradient approximation (GGA) with the Perdew-Burke-Ernzerhof (PBE)[32] exchange-correlation functional was used, combined with a double-ζ polarized (DZP) basis set. For this basis, each valence orbital is represented by two independent radial functions (double-ζ), along with additional polarization functions which are particularly optimized to accurately capture the directional character of the Mo d-orbitals and Se p-orbitals. This setup achieves an optimal balance between computational efficiency and accuracy for the study of two-dimensional $MoSe_2$. An energy shift of 250 meV was employed to control the cutoff radii of the pseudo-atomic orbitals, and a real-space mesh cutoff of 350 Ry was applied to ensure adequate convergence of the electron density. For static ground-state relaxations, k-point sampling was controlled through a k-grid cutoff of 15 Å, and spin-unpolarized calculations were performed because preliminary tests showed negligible impact of allowing for the possibility for spin to become polarized. Electron density mixing weight (0.001–0.1) and a density tolerance of $1\times10^{-3}$ were used to stabilize self-consistent field (SCF) convergence, and up to 60 SCF iterations were required to achieve convergence in most cases. Atomic positions were optimized with the conjugate-gradient method until residual forces converged below 0.04 eV/Å. Additionally, DFT-based molecular dynamics (DFT-MD) simulations were employed to examine the stability of $MoSe_2$ with different types of defects and surface terminations at multiple temperatures, thereby providing insights into the thermally induced structural changes and defect evolution. These simulations were performed using PBE/DZP, and a Nosé thermostat was adopted at 300 K, 650 K, and 700 K to capture the finite-temperature behavior. A density mixing weight of 0.1 was used, and up to 1000 SCF iterations were permitted to account for significant wavefunction changes that can arise during the course of a simulation. The MD simulations were carried out using a time step of 0.5 fs, which strikes a balance between accuracy in integrating the equations of motion and computational efficiency.

Based on DFT calculations, the cohesive energy of $MoSe_2$ was computed using

$$E_{coh} = E_{pristine} - n_{Mo}E_{Mo} - n_{Se}E_{Se} \qquad (1)$$

where $E_{pristine}$ is the energy of the compound, $E_{Mo}$ and $E_{Se}$ are the energies of an isolated Mo and Se atom, and $n_{Mo}$ and $n_{Se}$ are the numbers of corresponding atoms in the compound. The elastic constants were extracted from uniaxial stress–strain curves in the small-deformation regime. A fitting procedure was used to extract the polynomial of the finite-deformation Green tensor of different orders, and the second-order terms were used in the screening process. Vacancy formation energies were calculated with the following equation:

$$E_f = E_{defected} + n_{Mo}\mu_{Mo} + n_{Se}\mu_{Se} - E_{pristine} \qquad (2)$$

where $E_{defected}$ is the energy of the defected system, $E_{pristine}$ is the energy of the pristine system, $n_{Mo}$ and $n_{Se}$ are the number of missing Mo and Se atoms in the vacancy, and $\mu_{Mo}$ and $\mu_{Se}$ are chemical potentials for Mo and Se atoms, respectively.

In the transferability analysis, given the complexity of achieving a full description of the kinetics involved in formation and diffusion of multiple vacancies leading to inversion domains, we focused on elucidating the energetics of the inversion domain formation. To accomplish this, we performed Solid-State Nudged Elastic Band (SSNEB) simulations[33-35] using SIESTA and the Atomic Simulation Environment (ASE)[36] interface. A distinct feature of our approach was the explicit inclusion of stress evolution along the transition path, which we found to be particularly significant as the initial structure exhibited inherent stress while the final state was stress-free. A minimum energy path (MEP) was constructed using seven images between the initial and final states. The transition path was optimized using the climbing image modification of SSNEB, which ensures accurate identification of the transition state. Electronic structure calculations were performed using PBE/DZP with an energy shift of 0.001 Ry. We implemented a linear interpolation of stress tensors between the initially stressed and the final stress-free configurations across all intermediate images. The path optimization was performed using the FIRE algorithm[37] with carefully tuned parameters (maximum atomic displacement: 0.1 Å, time step: 0.1) to ensure stable convergence. The calculation was considered converged when the maximum force perpendicular to the path fell below 0.01 eV/Å. This approach allowed us to track both the structural transformation and the associated stress relaxation process simultaneously.

**Database preparation**

To generate a dataset representing various atomic environments for parameterizing ML potentials, we selected atomic configurations representing both equilibrium and non-equilibrium properties of $MoSe_2$. The corresponding energies and atomic forces for these configurations were obtained from DFT calculations as described above. For equilibrium properties, we included cohesive energy, the equation of state (near equilibrium), elastic constants ($C_{11}$ and $C_{12}$), and surface energies for armchair (AC) and zigzag (ZZ) surfaces. To capture non-equilibrium behavior, we incorporated uniaxial stress-strain curves/bond dissociation energy landscapes along the AC and ZZ directions, vacancy formation energies, phase transformation energies, and the stability of $MoSe_2$ with various defect types and surface terminations under different temperatures. To ensure robust training, we generated a diverse set of configurations spanning a wide range of strains and dissociation distances, covering the entire deformation process of MoSe$_2$. This approach enables the trained potential to accurately capture non-equilibrium large-deformation behaviors. The database was then divided into a training set and a test set, as shown in Fig. 1a. The training set was used for fitting and optimizing the ML potential, while the test set was used to assess the generalization performance and predictive accuracy of the trained potential.

**SNAP parameterization and optimization**

This section outlines the procedures used to train and optimize the SNAP potential by fitting DFT data. We selected SNAP as a representative descriptor-based ML potential due to its well-established accuracy in modeling interatomic interactions[22] and its implementation in the open-source Large-scale Atomic/Molecular Massively Parallel Simulator (LAMMPS), which facilitates efficient large-scale simulations. In the SNAP potential[22], the total energy of an atomic system $E$ is decomposed into a sum over atom energies $E_i$:

$$E = \sum_i E_i \qquad (3)$$

The atomic energy for atom $i$ is determined by its atomic environment (the positions and the species of neighboring atoms). The task of the machine-learning algorithms is to establish the relation between atomic energy and the configuration of the neighboring atoms. To apply machine learning algorithms, the atomic environment of each atom is represented by a descriptor vector that captures local structural information while maintaining rotational, translational, and permutational invariance. SNAP employs bi-spectrum components as its descriptors and the atomic energy is then expressed as a weighted summation of these bi-spectrum components:

$$E_i = \beta_{\eta i}^0 + \sum_{k=1}^{K} \beta_{\eta i}^k B_i^k \qquad (4)$$

where $B_i^k$ is the *k-th* bi-spectrum component, and $\beta_{\eta i}^k$ is the corresponding weight coefficient that depends on the element of atom $i$, $\eta_i$. The number of bi-spectrum coefficients for each element type is determined by $J_{max}$.

Based on the potential energy, one can also calculate the atomic force on atom $j$,

$$F_j = -\nabla_j E = -\nabla_j \sum_i E_i = -\sum_i \sum_{k=1}^{K} \beta_{\eta i}^k \nabla_j B_i^k \qquad (5)$$

For a given atomic system, where the atom positions are known, the values of $B_i^k$ and their derivatives are determined. Therefore, the energy and atomic forces are only the linear function of the coefficients $\beta_{\eta i}^k$. Equations (3) - (5) can be converted to matrix form Aβ = b, where β is a vector, whose elements are the SNAP coefficients to be determined, A is the matrix of bi-spectrum component, and b is a column vector whose elements are the observations (the potential energy and atomic forces) obtained from the DFT calculations. Usually, the number of unknown parameters is smaller than the number of data obtained from the DFT calculations, so the linear equation set can be solved as a least-squares problem. Assuming that P is a diagonal matrix whose elements are the weight factors associated with the observations, the problem becomes the minimization of the residue, $\varepsilon = |PA\beta - Pb|^2$. We randomly selected 10% of the configurations from the training set as subset 1 and left the remaining configurations as subset 2. As shown in Fig. 1b, this weighted least-squares problem was solved using the open-source software FitSNAP[38], where the SNAP coefficients were fitted to the training subset 2. It is noted that in addition to SNAP coefficients, some hyperparameters, including cutoff radius $R_{cut}$, and weight $w$ for each element, need to be optimized when determining the coefficients. Properly optimized hyperparameters enhance the accuracy of the potential. In this study, we optimized the SNAP potential hyperparameters using Bayesian optimization, employing the energy and force errors of training subset 1 as optimization metrics. The hyperparameters also include $J_{max}$, which defines the descriptor length. To examine the trade-off between accuracy and computational cost, we evaluated $J_{max}$ at values of 2, 3, 4, and 5. For each $J_{max}$ value, we optimized the SNAP potential using the previously outlined process.

**Allegro parameterization and optimization**
The Allegro potential uses a deep equivariant neural network architecture that has been successfully demonstrated for a wide variety of molecular and material systems[23]. Compared to its equivariant graph neural network predecessors[39-41], Allegro retains state-of-the-art accuracy[23] while exhibiting enhanced scalability in both time to solution and simulation size[42]. This is accomplished through a local description of the atomic environment as well as an efficient network design focusing most of the model capacity on faster scalar operations with minimal interactions with more computationally intensive tensor operations[14]. Using a grid search approach, we select optimal values for hyperparameters including the radial cutoff for pairwise interactions ($r_c$), the degree of the polynomial cutoff function ($p$), and the multiplicity of the embedded environment ($n_{feats}$). By evaluating the different combinations of these hyperparameters on a subset of the training data, we chose the following values: $r_c = 6\text{Å}$, $p = 12$, and $n_{feats} = 64$. In terms of model capacity, we chose a single Allegro layer, the 2-body latent multilayer perceptron (MLP) has three hidden layers with dimensions [32,64,128], and the later latent MLP also has three layers with dimensions [128,128,128] (both MLPs use SiLU nonlinearities on the outputs of the hidden layers). The embedding weight projection is implemented as a single matrix multiplication without any hidden layers or nonlinearities. Similarly, the final edge energy MLP includes one hidden layer of dimension 128 and does not use nonlinearity. We choose a mean squared error (MSE) loss function with the forces receiving a weight of 100, the per-atom energies receiving a weight of 1, and the stresses receiving a weight of $1\times10^{-6}$. The model is trained with a batch size of 1 using an initial learning rate of 0.001. The learning rate was reduced using an on-plateau scheduler based on the validation loss with a patience of 30 and a decay factor of 0.75. Training was stopped when either the learning rate dropped below $1\times10^{-5}$, validation loss did not improve for 30 epochs, or a maximum of 10,000 epochs were completed.

Like the SNAP potential, in Allegro the energy of a system is decomposed as a sum of atomic energies however it includes per-species scaling ($\sigma_{Z_i}$, default: $F_{rms}$ across entire dataset) and shift parameters ($\mu_{Z_i}$, default: mean $E_{coh,i}$):

$$E = \sum_i^N \sigma_{Z_i} E_i + \mu_{Z_i} \quad (6)$$

The atomic energies are further decomposed into sum of pairwise energies with a per-species-pair scaling parameter ($\sigma_{Z_i,Z_j}$, default: $F_{rms}$ across entire dataset):

$$E_i = \sum_{j \in N(i)} \sigma_{Z_i,Z_j} E_{ij} \quad (7)$$

Finally, the forces acting on atom $i$ are computed using autodifferentiation:

$$\vec{F_i} = -\nabla_i E \quad (8)$$

The trainable parameters of the model are optimized by minimizing the following mean squared error (MSE) loss function (per-atom MSE for energy term):

$$\mathcal{L} = \frac{w_E}{B}\sum_b^B \left(\frac{\hat{E}_b - E_b}{N}\right)^2 + \frac{w_F}{3BN}\sum_{i=1}^{BN}\sum_{\alpha=1}^3 \left\| -\frac{\partial \hat{E}}{\partial r_{i,\alpha}} - F_{i,\alpha} \right\|^2 \quad (9)$$

where $B$ is the batch size, $N$ is the number of atoms in the system, $E_b$ is the batch of true energies, $\hat{E}_b$ is the batch of predicted energies, $F_{i,\alpha}$ is the force component on atom $i$ in spatial direction $\alpha$, $w_E$ is the energy weight and $w_F$ is the force weight.

## Acknowledgements

H.D.E. acknowledges financial support from the Office of Naval Research (grant N000142212133) and the National Science Foundation (grant CMMI-1953806). This work was partially supported by the US National Science Foundation, Award 2124241 and Purdue University Rosen Center for Advanced Computing. Computational resources provided by the Center of Nanoscale Materials at Argonne National Laboratory and the Quest High Performance Computing Cluster at Northwestern University are acknowledged. M.J.M. gratefully acknowledges support from the Laboratory Directed Research and Development program at Sandia National Laboratories through the Materials Science Research Foundation. Sandia National Laboratories is a multi-mission laboratory managed and operated by National Technology & Engineering Solutions of Sandia, LLC (NTESS), a wholly owned subsidiary of Honeywell International Inc., for the U.S. Department of Energy's National Nuclear Security Administration (DOE/NNSA) under contract DE-NA0003525. This written work is authored by an employee of NTESS. The employee, not NTESS, owns the right, title and interest in and to the written work and is responsible for its contents. Any subjective views or opinions that might be expressed in the written work do not necessarily represent the views of the U.S. Government. The publisher acknowledges that the U.S. Government retains a non-exclusive, paid-up, irrevocable, world-wide license to publish or reproduce the published form of this written work or allow others to do so, for U.S. Government purposes. The DOE will provide public access to results of federally sponsored research in accordance with the DOE Public Access Plan. The authors also acknowledge technical discussions about DFT models and calculations of the J-integral with Dr. Hoang Nguyen.

## Contributions

**Y. Z.** contributed to modeling, data analysis, and writing the manuscript. **R.J.A.** contributed to modeling, data analysis, and writing the manuscript. **K.L.** contributed to DFT calculations and writing the manuscript. **M.J.M** contributed to modeling with SNAP. **J.T.P.** contributed to DFT calculations. **S.K.R.S.K.** contributed to facilities and manuscript editing. **A.S.** contributed to conceptualization, modeling, data analysis, funding acquisition, and manuscript editing. **H.D.E.** contributed to conceptualization, modeling, data analysis, funding acquisition, and manuscript editing.

# Supplementary Information

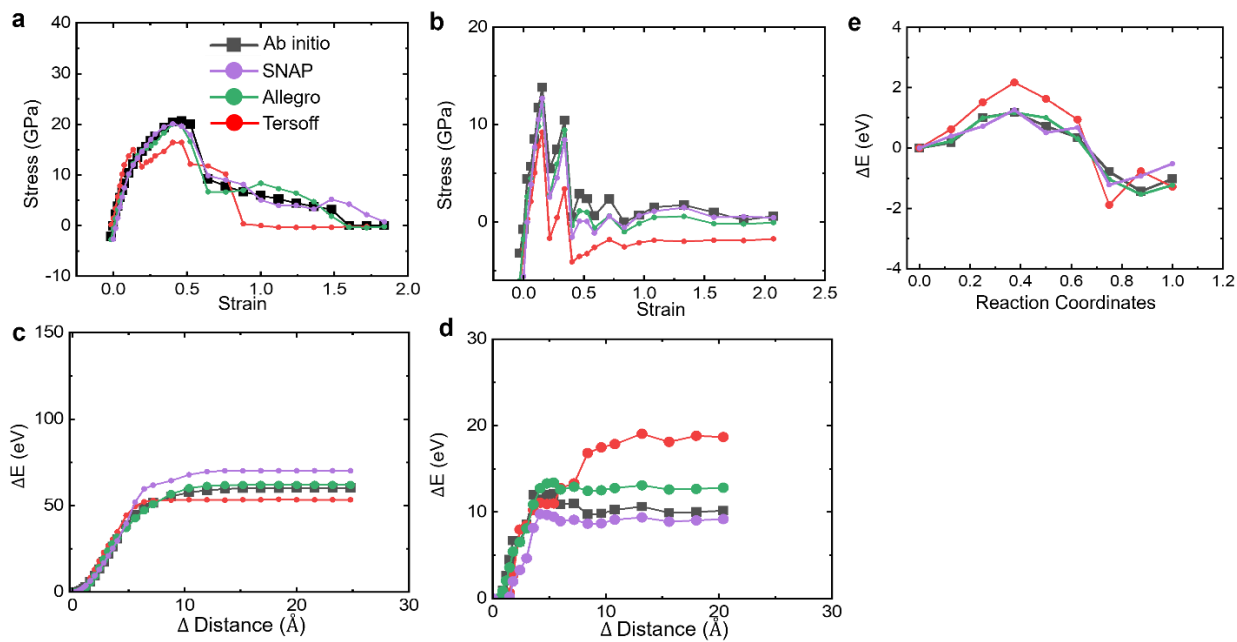

**Fig. S1.** Predictions of the properties in training set by parameterized potentials in comparison to ab initio results. (a) Uniaxial stress-strain relation for $MoSe_2$ along the armchair direction. (b) Nonuniform stress-strain relation for $MoSe_2$ with 4|4E grain boundaries along the zigzag direction. (c) Bond dissociation energy landscape along the zigzag direction. (d) Nonuniform bond dissociation energy landscape for $MoSe_2$ with 4|4E grain boundaries along the zigzag direction. (e) 4B4G phase transition energy landscape.

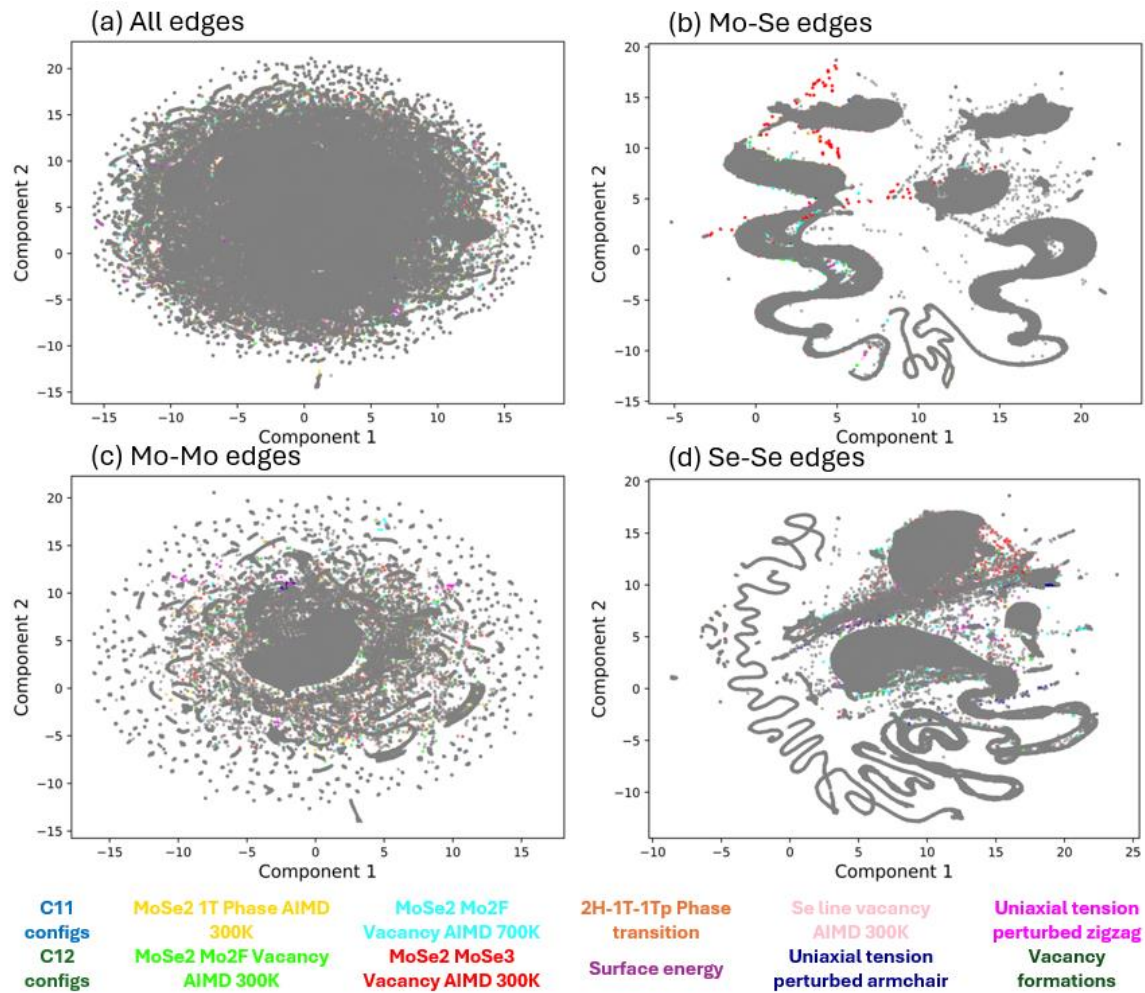

**Fig. S2.** Visualization of the UMAP reduced edge features for all edges (a), Mo-Se edges (b), Mo-Mo edges (c), and Se-Se edges (d). The configurations from the training set are plotted in grey and the configurations from the test set are colored corresponding to the type of simulation.

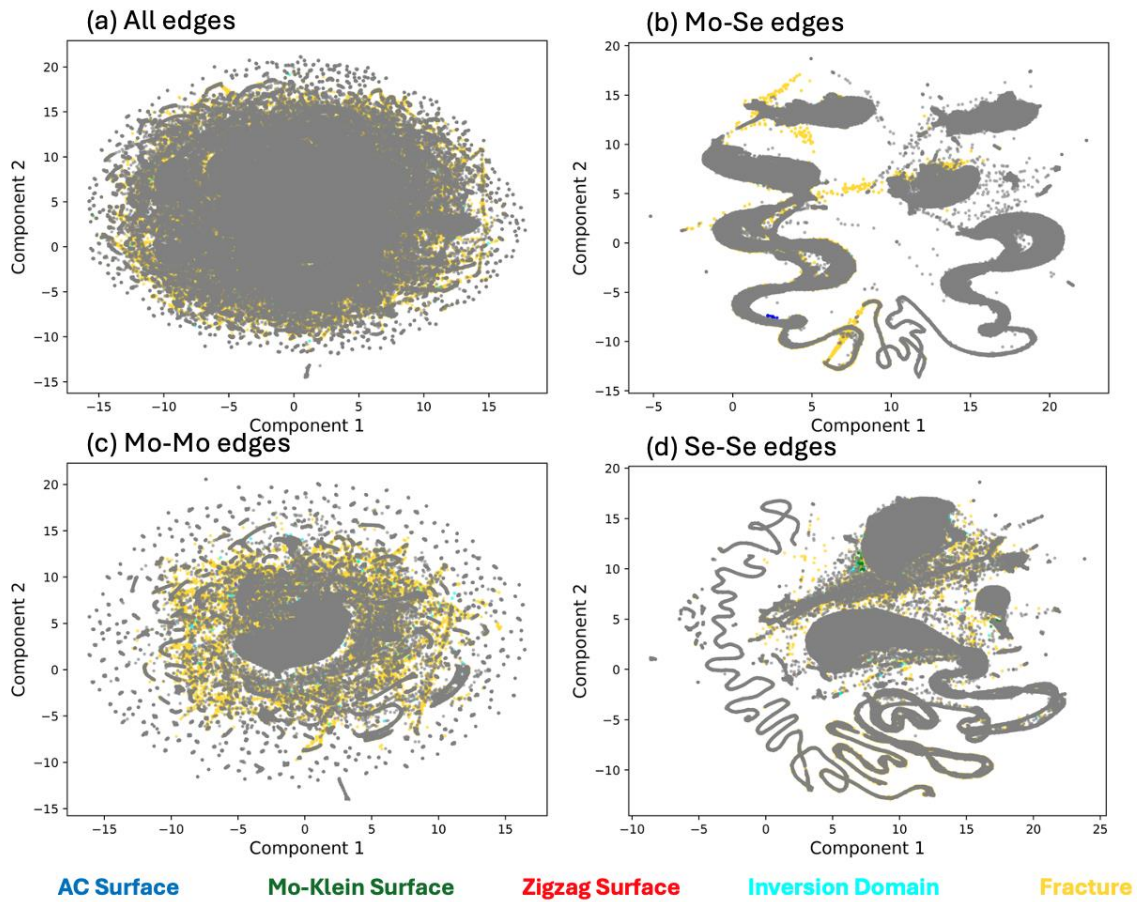

**Fig. S3.** Visualization of the UMAP reduced edge features for all edges (a), Mo-Se edges (b), Mo-Mo edges (c), and Se-Se edges (d). The configurations from the training set are plotted in grey and the configurations from the transferability set are colored corresponding to the type of simulation.

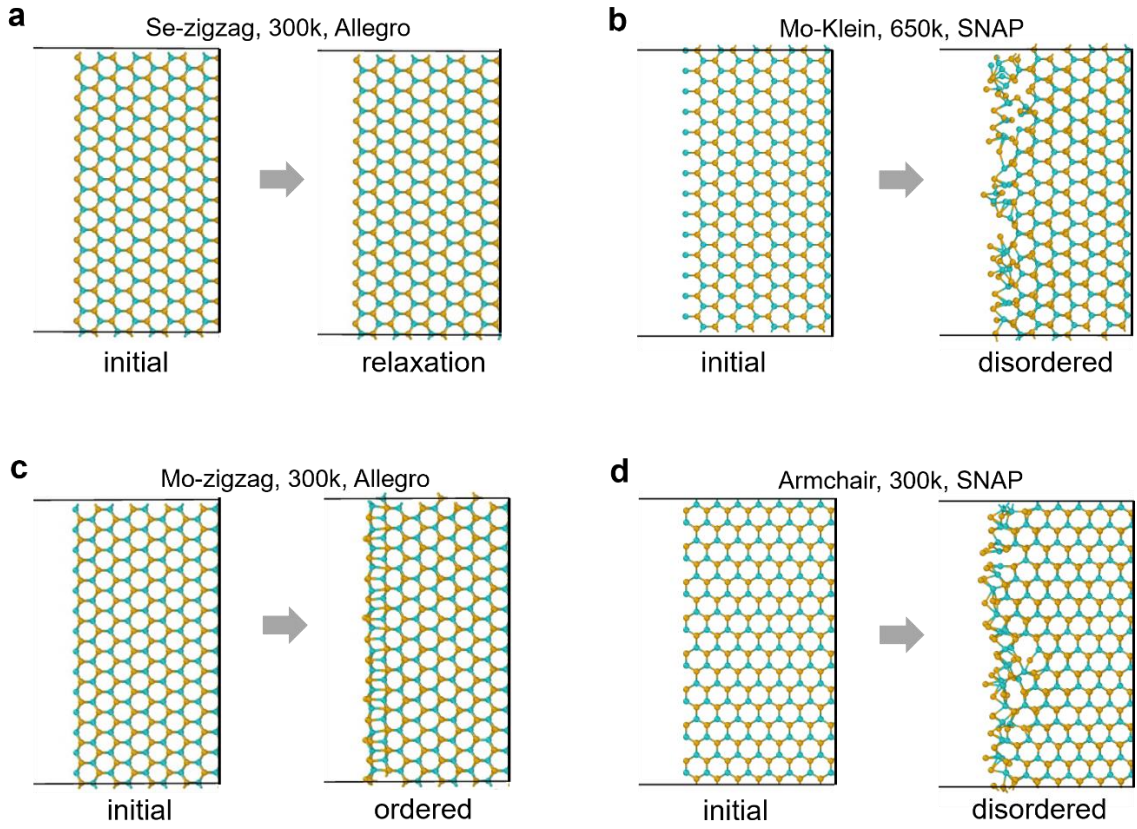

**Fig. S4.** Initial and final atomic structures of different surface edge types in MD simulations. (a) Se-zigzag surface edge type at 300K. (b) Mo-Klein surface edge type at 650K. (c) Mo-zigzag surface edge type at 300K. (d) Armchair surface edge type at 300K.

|  | $\Delta E = E_{reconstructed} - E_{initial}$ (eV) |
| --- | --- |
| DFT | -1.5 |
| Allegro | -1.5 |
| SNAP | 2.9 |
| Tersoff | 18.5 |

**Table. S1.** Energy differences between initial and reconstructed atomic structures of the Mo-zigzag surface edge type at 300K, predicted by parameterized potentials in comparison to DFT result.

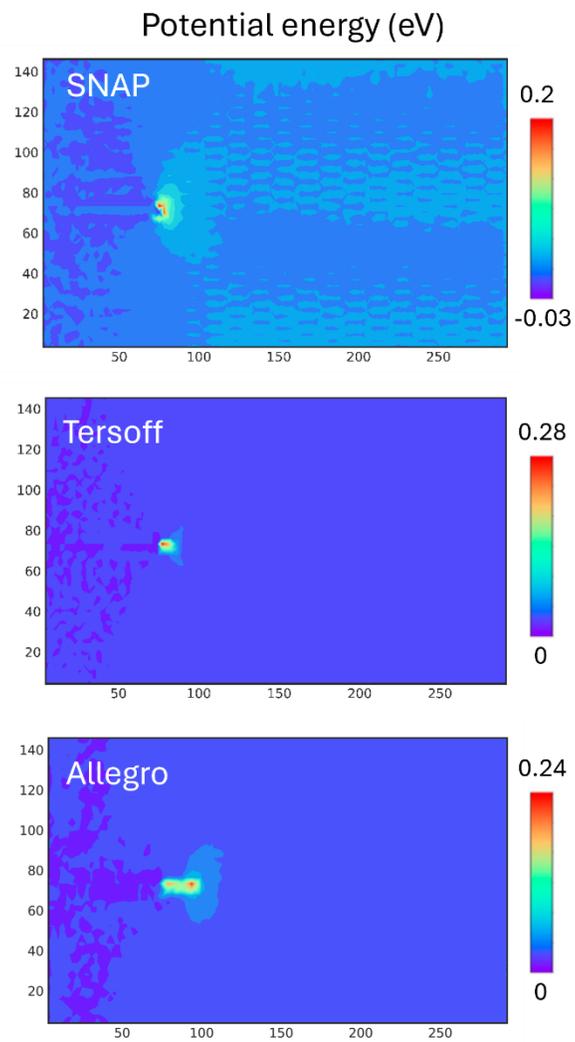

**Fig. S5.** Potential energy in MoSe$_2$ calculated with SNAP, Tersoff and Allegro potentials.

|          | Experiment | Allegro ($l_{max}=3$) | Allegro ($l_{max}=2$) | Allegro ($l_{max}=1$) |
|----------|------------|-----------------------|-----------------------|-----------------------|
| J (N/m)  | 3.20       | 3.29                  | 3.24                  | 3.54                  |

|          | SNAP ($J_{max}=5$) | SNAP ($J_{max}=4$) | SNAP ($J_{max}=3$) | SNAP ($J_{max}=2$) |
|----------|--------------------|--------------------|--------------------|--------------------|
| J (N/m)  | 3.42               | 3.49               | 2.63               | 2.17               |

**Table. S2.** Fracture toughness calculated using the SNAP potential with different $J_{max}$ values and the Allegro potential with different $l_{max}$ values.